\def\rfr#1{eq. (\ref{#1})}

\def\leti{Lense--Thirring}

\def\bar{\begin{eqnarray}}
\def\ear{\end{eqnarray}}

\def\eqi{\begin{equation}}
\def\eqf{\end{equation}}
\def\eqia{\begin{eqnarray}}
\def\eqfa{\end{eqnarray}}
\def\rp#1#2{{#1\over#2}}

\def\lb#1{\label{#1}}

\def\oc2{$\mathcal{O}(c^{-2})$}

\documentclass[11pt]{article}
\usepackage{amsmath,amsthm,amscd,amssymb}
\usepackage{latexsym}
\usepackage{graphicx,epsfig}

\begin{document}

\noindent{\bf \LARGE{The impact of the new Earth gravity models on
the measurement of the Lense-Thirring effect with a new satellite
}}
\\
\\
\\
{Lorenzo Iorio}\\
{\it Dipartimento Interateneo di Fisica dell' Universit${\rm
\grave{a}}$ di Bari
\\Via Amendola 173, 70126\\Bari, Italy
\\e-mail: lorenzo.iorio@libero.it}

\begin{abstract}
In this paper we investigate the opportunities offered by  the new
Earth gravity models from the dedicated CHAMP and, especially,
GRACE missions to the project of measuring the general
relativistic Lense-Thirring effect with a new Earth's artificial
satellite. It turns out that it would be possible to abandon the
stringent, and expensive, requirements on the orbital geometry of
the originally prosed LARES mission (same semimajor axis $a=12270$
km of the existing LAGEOS and inclination $i=70$ deg) by inserting
the new spacecraft in a relatively low, and cheaper, orbit
($a=7500-8000$ km, $i\sim 70$ deg) and suitably combining its node
$\Omega$ with those of LAGEOS and LAGEOS II in order to cancel out
the first even zonal harmonic coefficients of the multipolar
expansion of the terrestrial gravitational potential $J_2$, $J_4$
along with their temporal variations. The total systematic error
due to the mismodelling in the remaining even zonal harmonics
would amount to $\sim 1\%$ and would be insensitive to  departures
of the inclination from the originally proposed value of many
degrees. No semisecular long-period perturbations would be
introduced because the period of the node, which is also the
period of the solar $K_1$ tidal perturbation, would amount to
$\sim 10^2$ days. Since the coefficient of the node of the new
satellite would be smaller than 0.1 for such low altitudes, the
impact of the non-gravitational perturbations of it on the
proposed combination would be negligible. Then, a particular
financial and technological effort for suitably building the
satellite in order to minimize the non-conservative accelerations
would be unnecessary.
\end{abstract}
\newpage
\tableofcontents
\newpage
\section{Introduction}
The general relativistic gravitomagnetic Lense-Thirring effect on
the orbit of a test particle (Lense and Thirring 1918) consists of
secular precessions of the longitude of the ascending node
$\Omega$ and the argument of the pericentre $\omega$ of the orbit
of a test particle in geodesic motion around a central spinning
mass with proper angular momentum $J$ \eqi\dot\Omega_{\rm
LT}=\rp{2GJ}{c^2 a^3(1-e^2)^{3/2}},\ \dot\omega_{\rm
LT}=-\rp{6GJ\cos i}{c^2 a^3(1-e^2)^{3/2}},\eqf where $G$ is the
Newtonian gravitational constant and $a,i,e$ are the semimajor
axis, the inclination to the equatorial plane of the central mass
and the eccentricity, respectively, of the test particle's orbit.

Up to now, a clear and undisputable direct test of such a general
relativistic prediction, which may have important consequences in
many astrophysical scenarios involving, e.g., accreting disks
around black holes (Stella et al. 2003), is not yet available.
However, according to K. Nordtvedt gravitomagnetism would have
already been indirectly tested  from the radial motion of the
LAGEOS Earth satellite (Nordtvedt 1988) and from the
high-precision reconstruction of the lunar orbit with the Lunar
Laser Ranging (LLR) technique (Nordtvedt 2003).

In April 2004 the GP-B mission (Everitt et al. 2001) has been
launched. Its aim is the measurement of another gravitomagnetic
effect, i.e. the precession of the spins (Schiff 1960) of four
superconducting gyroscopes carried onboard with a claimed accuracy
of $1\%$ or better.
\subsection{The performed tests with the existing LAGEOS and LAGEOS II satellites}
Up to now, the only performed tests aimed at the explicit and
direct detection of such a post-Newtonian effect in the
gravitational field of the Earth have been performed by Ciufolini
and coworkers by analyzing the data from the existing Satellite
Laser Ranging (SLR) LAGEOS and LAGEOS II satellites (Ciufolini et
al. 1998, Ciufolini and Pavlis 2004). The claimed total accuracy
is $20-30\%$ (1-sigma) and, for the most recent test, $5-10\%$
(1-3 sigma). However, the evaluation of the total error budget,
including the impact of various sources of systematic errors
induced by many gravitational (static and time-dependent parts of
the Earth's gravitational potential; Iorio 2001, 2003d) and
non-gravitational (direct solar radiation pressure, Earth's
albedo, Earth's infrared radiation, asymmetric reflectivity, solar
Yarkovsky-Schach and Earth's Yarkovsky-Rubincam thermal thrusts;
Lucchesi 2001, 2002, 2003, 2004, Lucchesi e al. 2004)
perturbations has been critically discussed by a number of authors
(Ries et al. 2003, Vespe and Rutigliano 2004, Iorio 2005a) who
pointed out that such evaluations are to be considered optimistic.

This is particularly true for the tests involving the nodes of
LAGEOS and LAGEOS II and the perigee of LAGEOS II (Ciufolini et
al. 1998) based on a suitable linear combination of them proposed
in (Ciufolini 1996). Indeed, in that case the total error might be
close to 100$\%$ (1-sigma) mainly due to the impact of the
non--gravitational perturbations, which severely affect the
perigee of the LAGEOS-like satellites, and of the mismodelling in
the static part of the even zonal harmonic coefficients $J_{\ell}$
of the multipolar expansion of the terrestrial gravitational
potential. Indeed, according to the pre-CHAMP/GRACE era EGM96
Earth gravity model (Lemoine et al. 1998) adopted in that
analysis, the systematic error due to the even zonal harmonics
would be $\sim 80\%$ (Ries et al. 2003, Vespe and Rutigliano 2004,
Iorio 2003d, 2005a) at 1-sigma.

In the case of the more precise test involving only the nodes of
the LAGEOS satellites (Ciufolini and Pavlis 2004), based on a
suitable linear combination of them proposed in (Iorio and Morea
2004) and the EIGEN-GRACE02S Earth gravity model (Reigber et al.
2005), the impact of the secular variations $\dot J_{\ell}$ of the
low-degree even zonal harmonics of the geopotential may be a
further biasing effect. According to (Iorio 2005a), the total
error could be $\sim 15-45\%$ at 1-3 sigma over the 11-year time
span of the performed analysis.

A possible way to solve such problems has recently been proposed
in (Iorio and Doornbos 2005). It consists in suitably combining
the nodes of the existing geodetic LAGEOS, LAGEOS II and Ajisai
satellites and of the altimeter Jason-1 satellite in order to
cancel out $J_2, J_4, J_6$ along with their secular variations.
The systematic error due to the remaining even zonal harmonics is
$< 2\%$ according to the recently released EIGEN-CG01C Earth
gravity model  (Reigber et al. 2004) which combines data from
CHAMP, GRACE and terrestrial gravimetry and altimetry
measurements. However, the reduction of the data of Ajisai and,
especially, Jason-1 in a truly dynamical way to a level comparable
to that of the LAGEOS satellites would not be a trivial task,
mainly due to the action of the non-gravitational perturbations.
However, it must be pointed out that the coefficients with which
Ajisai and Jason-1 enter this combination are of the order of
$10^{-3}-10^{-2}$ and no very long period sinusoidal perturbations
are introduced. This feature is very important because it would,
thus, be possible to fit and remove them from the signal over a
not too long observational time span.
\subsection{The proposed LAGEOS III/LARES/WEBER-SAT satellite}
In 1986 (Ciufolini 1986),  Ciufolini proposed to orbit a
LAGEOS-like satellite--known also as LAGEOS III (Tapley and
Ciufolini 1989), LARES (Ciufolini 1998), WEBER-SAT (Ciufolini et
al. 2004)--with the same orbital parameters of LAGEOS, apart from
the inclination which should have an offset of 180 deg, in order
to cancel out the impact of all the even zonal harmonics of the
geopotential, at least in principle. Indeed, while the nodal
Lense-Thirring precessions are independent of the inclination of
the orbital planes, the classical nodal precessions are
proportional to $\cos i$, so that, by using the sum of the nodes
of LAGEOS and its twin, the gravitomagnetic rates would  sum up
while the aliasing classical rates would subtracted.

In Table \ref{para} we quote the orbital parameters of the
existing LAGEOS satellites and of the proposed LARES.
{\small\begin{table}\caption{Orbital parameters of the existing
LAGEOS and LAGEOS II and of the proposed LARES and their
Lense-Thirring node precessions in milliarcseconds per year (mas
yr$^{-1}$ ). }\label{para}

\begin{tabular}{lllll}
\noalign{\hrule height 1.5pt}

Satellite & $a$ (km) & $e$ & $i$ (deg) & $\dot\Omega_{\rm LT}$ (mas yr$^{-1}$)  \\

\hline

LAGEOS    &  12270    & 0.0045 &  110 & 31\\
LAGEOS II &  12163    & 0.0135 & 52.64 & 31.5\\
LARES    & 12270    & 0.04  & 70 & 31\\

\noalign{\hrule height 1.5pt}
\end{tabular}

\end{table}}
\subsection{Aim of the paper}
In this paper we wish to investigate the impact that the new Earth
gravity models from the dedicated CHAMP
(http://www.gfz-potsdam.de/pb1/op/champ/index$\_$CHAMP.html) and
GRACE (http://www.gfz-potsdam.de/pb1/op/grace/index$\_$GRACE.html)
missions might have on a space mission involving the launch of
only one new spacecraft (Section \ref{una}) or two new spacecraft
(Section \ref{dua}) aimed at the measurement of the Lense-Thirring
effect. In Table \ref{geomodels} we show the calibrated sigmas of
the even zonal harmonic coefficients of the geopotential of the
pre-CHAMP/GRACE EGM96 Earth gravity model and of some of the most
recent GRACE/CHAMP-based solutions EIGEN-GRACE02S and EIGEN-CG01C,
released by the GeoForschungsZentrum (GFZ), Potsdam, and GGM02S
(http://www.csr.utexas.edu/grace/gravity/), released by the Centre
for Space research (CSR), Austin.
\begin{table}[ht!]
\caption{Errors in the even zonal normalized Stokes coefficients
$\sigma_{\overline{C}_{\ell 0}}$ for various Earth gravity models
up to degree $\ell=20$.  It is important to note that, while EGM96
is based on a multidecadal time span, the CHAMP and GRACE models
presented here have been obtained from observational intervals of
a few months: 111 days from GRACE for EIGEN-GRACE02S, 860 days
from CHAMP and 200 days from GRACE (plus terrestrial gravimetry
and altimetry data) for EIGEN-CG01C, 363 days from GRACE for
GGM02S. } \label{geomodels}
\begin{center}
\begin{tabular}{lllll}
\noalign{\hrule height 1.5pt}
$\ell$ & EGM96 & EIGEN-GRACE02S & EIGEN-CG01C & GGM02S\\
\hline 2 & 3.561$\times 10^{-11}$ & $5.304\times 10^{-11}$ & $3.750\times 10^{-11}$ & $1.1\times 10^{-10}$\\
4 & 1.042$\times 10^{-10}$ & $3.921\times 10^{-12}$ & $6.242\times 10^{-12}$ & $8.3\times 10^{-12}$\\
6 & 1.449$\times 10^{-10}$ & $2.049\times 10^{-12}$ & $2.820\times 10^{-12}$ & $4.5\times 10^{-12}$\\
8 & 2.266$\times 10^{-10}$ & $1.479\times 10^{-12}$ & $1.792\times 10^{-12}$ & $2.8\times 10^{-12}$\\
10 & 3.089$\times 10^{-10}$ & $2.101\times 10^{-12}$ & $1.317\times 10^{-12}$& $2\times 10^{-12}$\\
12 & 4.358$\times 10^{-10}$ & $1.228\times 10^{-12}$ & $1.053\times 10^{-12}$& $1.8\times 10^{-12}$\\
14 & 5.459$\times 10^{-10}$ & $1.202\times 10^{-12}$ & $8.931\times 10^{-13}$& $1.6\times 10^{-12}$\\
16 & 5.313$\times 10^{-10}$ & $9.945\times 10^{-13}$ & $7.905\times 10^{-13}$& $1.6\times 10^{-12}$\\
18 & 4.678$\times 10^{-10}$ & $9.984\times 10^{-13}$ & $7.236\times 10^{-13}$& $1.6\times 10^{-12}$\\
20 & 4.690$\times 10^{-10}$ & $1.081\times 10^{-12}$ & $6.784\times 10^{-13}$& $1.6\times 10^{-12}$\\
\noalign{\hrule height 1.5pt}
\end{tabular}
\end{center}
\end{table}
We will show that, with only one new satellite, its data should be
combined with those of the existing LAGEOS  and LAGEOS II. While
the gravitational error could be pushed below the $1\%$ level by
the forthcoming, more robust solutions for the terrestrial
gravitational field, the error due to the non-gravitational
perturbations would remain $\sim 1\%$ due to their impact on the
LAGEOS satellites. However, the implementation of such a mission
would be much facilitated by the fact that  an orbit lower than
that of LAGEOS, as in the original LAGEOS-LARES experiment, would
be acceptable for the new laser target.

Only the launch of at least two new spacecraft endowed with some
active mechanism of compensation of the non-gravitational
perturbations could allow to discard the passive LAGEOS satellites
and, then, to reduce the total systematic error below the $1\%$
threshold fully exploiting the benefits from the CHAMP and GRACE
models. Of course, this option would be, in practice, much less
feasible because of its higher cost.

\section{What can be done with only one new, passive satellite of LAGEOS
type?}\lb{una} In this Section we will analyze the impact of the
new Earth gravity model solutions from the dedicated CHAMP and
GRACE missions on a scenario in which only one new passive
satellite of LAGEOS type is at our disposal. We will show that it
is not possible to only use its data. Instead, a suitable
combination of them with those of the existing LAGEOS satellites
would allow to obtain a robust and reliable measurement of the
Lense-Thirring effect at the $\sim 1\%$ level with a lower-cost
mission with respect to the originally proposed LARES orbital
configuration. Indeed, in regard to the semimajor axis, crucial
for the cost of the mission whose most expensive part is
represented by the launcher, it would be possible to pass from
$12270$ km to $7-8000$ km only. The Earth gravity model
EIGEN-CG01C has been adopted.
\subsection{The originally proposed LAGEOS-LARES scenario: the sum of the nodes} The
originally proposed LAGEOS-LARES/LAGEOS III mission (Ciufolini
1986) can only be realized if the LARES satellite has its orbital
parameters as closest as possible to those of the existing LAGEOS,
with the inclination 180 deg apart. Such strict constraints would
make the launch of LARES more demanding in terms of quality, power
and cost of the launcher. This could become a serious drawback for
a specifically dedicated space mission which could improve the
present-day accuracy in knowing the Lense-Thirring effect of more
or less one order of magnitude with respect to the already
performed or proposed tests with the existing satellites. Indeed,
the use of the data from the old-generation passive LAGEOS
satellite would keep the systematic error due to the
non-gravitational perturbations at the 1-2$\%$ level,
irrespectively of what the improvement in the systematic error of
gravitational origin could be.

Another important point is that the simple sum of the nodes of two
satellites, by construction, does not cancel out any even zonal
harmonic: an exact cancellation occurs only if the semimajor axis
and the eccentricities are identical and the inclinations are 180
deg apart. Since this cannot be realized in the real world, the
sum of the nodes remains exposed to the residual corrupting action
of all the even zonal harmonics of the geopotential and of their
secular variations. This topic has been addressed in (Iorio 2003a)
with the EGM96 model.

\begin{figure}
\begin{center}
\includegraphics[width=15cm,height=12cm]{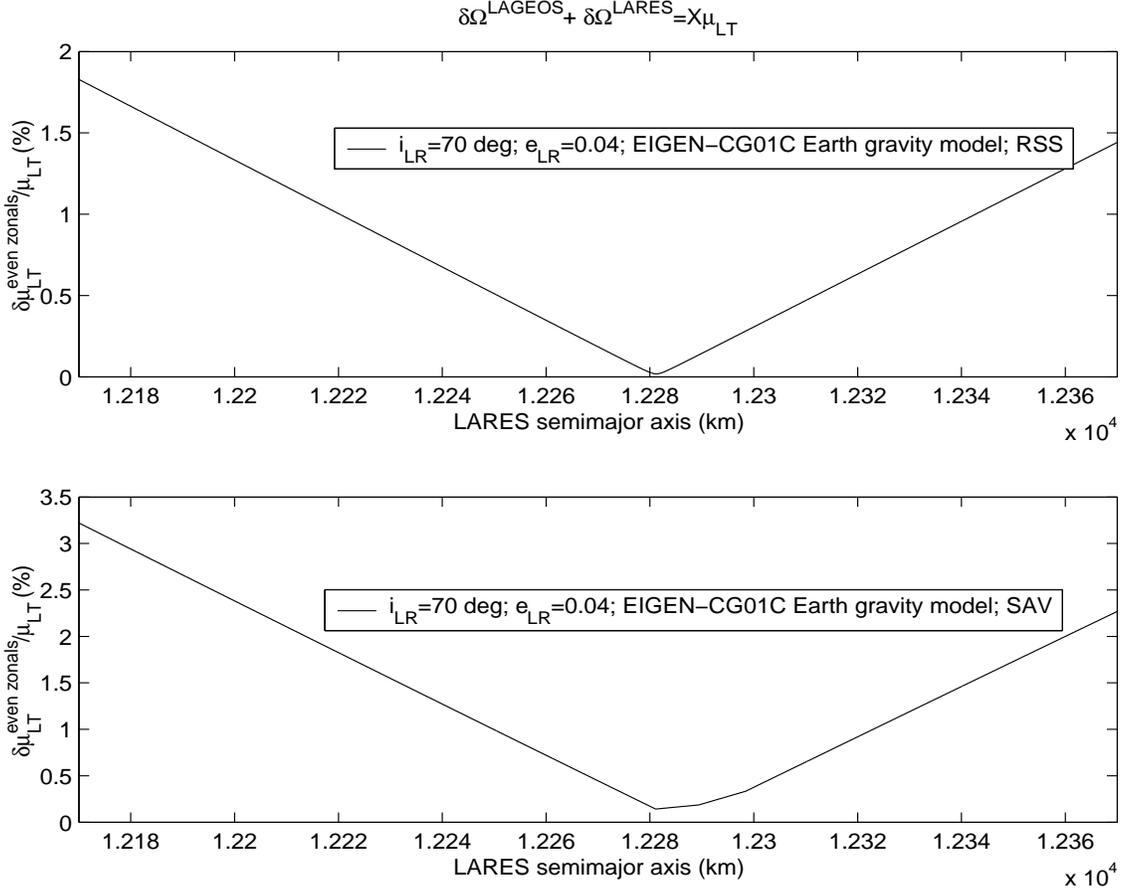}
\end{center}
\caption{\label{CG01C_LARES_a} Dependence on $a$ of the percent
1-sigma systematic error $\delta\mu^{\rm even\ zonals}_{\rm
LT}/\mu_{\rm LT}$ due to the mismodelling in the static part of
the even zonal harmonics of the geopotential according to
EIGEN-CG01C for the sum of the residuals of the nodes
$\delta\Omega^{\rm LAGEOS}+\delta\Omega^{\rm LARES}$ of the
originally proposed LAGEOS-LARES configuration with $i=70$ deg km
and $e=0.04$. The secular variations of the even zonal harmonics
$\dot J_2,\dot J_4,\dot J_6$ have been included. The chosen time
span is 10 years. In the upper panel the Root-Sum-Square (RSS)
calculation is presented, while in the lower panel the linear Sum
of the Absolute Values (SAV) of the individual errors is shown.}
\end{figure}

In Figure \ref{CG01C_LARES_a} we show the impact of the
unavoidable orbital injection errors in the semimajor axis
according to the EIGEN-CG01C model.
It can be noted that, despite the improvements in our knowledge of
the terrestrial gravity field, the orbital height of LARES should
still be similar to that of LAGEOS.

\begin{figure}
\begin{center}
\includegraphics[width=15cm,height=12cm]{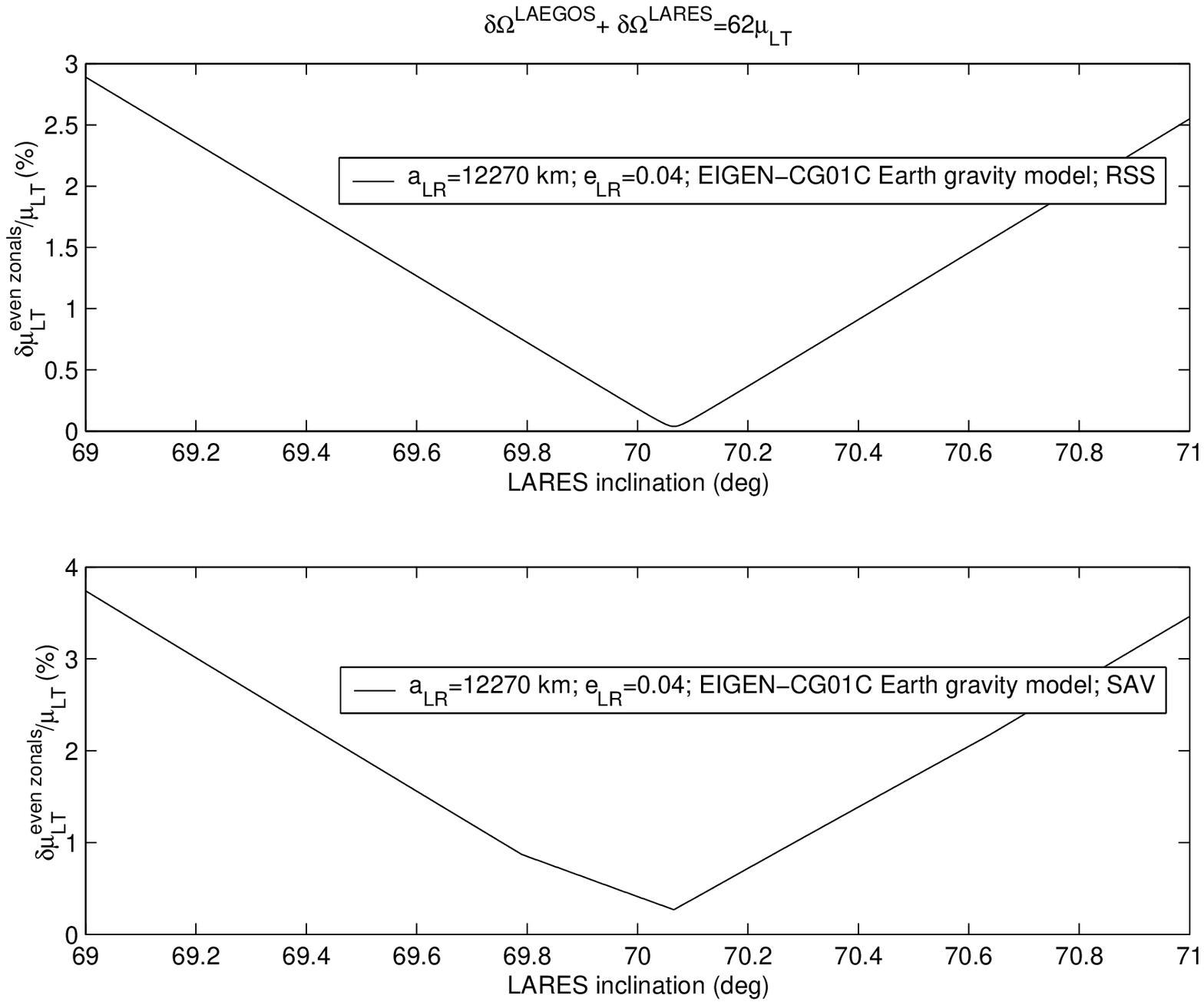}
\end{center}
\caption{\label{CG01C_LARES_i} Dependence on $i$ of the percent
1-sigma systematic error $\delta\mu^{\rm even\ zonals}_{\rm
LT}/\mu_{\rm LT}$ due to the mismodelling in the static part of
the even zonal harmonics of the geopotential according to
EIGEN-CG01C for sum of the residuals of the nodes
$\delta\Omega^{\rm LAGEOS}+\delta\Omega^{\rm LARES}$ of the
originally proposed LAGEOS-LARES configuration with $a=12270$ km
and $e=0.04$. The secular variations of the even zonal harmonics
$\dot J_2,\dot J_4,\dot J_6$ have  been included. The chosen time
span is 10 years. In the upper panel the Root-Sum-Square (RSS)
calculation is presented, while in the lower panel the linear Sum
of the Absolute Values (SAV) of the individual errors is shown.}
\end{figure}

In Figure \ref{CG01C_LARES_i} the impact of the departures of the
inclination from its nominal value is shown.
Again, the need for a high accuracy in inserting LARES in orbit
with an inclination as closest as possible to its nominal value is
apparent.

It must be pointed out that in Figure \ref{CG01C_LARES_a} and
Figure \ref{CG01C_LARES_i} also the impact of the secular
variations of the low-degree even zonal harmonics $\dot J_2,\dot
J_4,\dot J_6$ has been included and calculated over 10 years.
Since their integrated shift grows quadratically in time,  their
corrupting impact would become relatively important over
observational time spans some years long due to the orbital
injection errors.
\subsubsection{Conclusions}
In conclusion, the originally proposed LAGEOS-LARES scenario could
work only with a relatively expensive mission mainly due to the
need of a launcher of high performance and quality in order to
reduce the orbital injection errors. The total accuracy would
amount to a few percent and would depend on the precision with
which the nominal orbit could be realized in practice. Moreover,
the secular variations of all the low-degree even zonal harmonics
would further corrupt the recovery of the Lense-Thirring
precessions over observational time spans some years long. In view
of the already performed measurements and of the suggested
alternative tests with the existing satellites the originally
proposed LAGEOS-LARES two-nodes scenario appears now difficult to
be supported.
\subsection{The LARES-only scenario} In this Section we will
examine the possibility of only using the node of LARES in order
to exploit its reduced sensitivity to the non-gravitational
perturbations thanks to its proposed particular construction. More
precisely, we want to see if discarding the LAGEOS data could
allow to push the total systematic error below the 1$\%$ level.
The EIGEN-CG01C Earth gravity model is adopted in order to assess
the impact of the classical part of the terrestrial gravitational
field.  Unfortunately, the answer is negative.
\subsubsection{The nearly-polar orbit}
\begin{figure}
\begin{center}
\includegraphics[width=15cm,height=12cm]{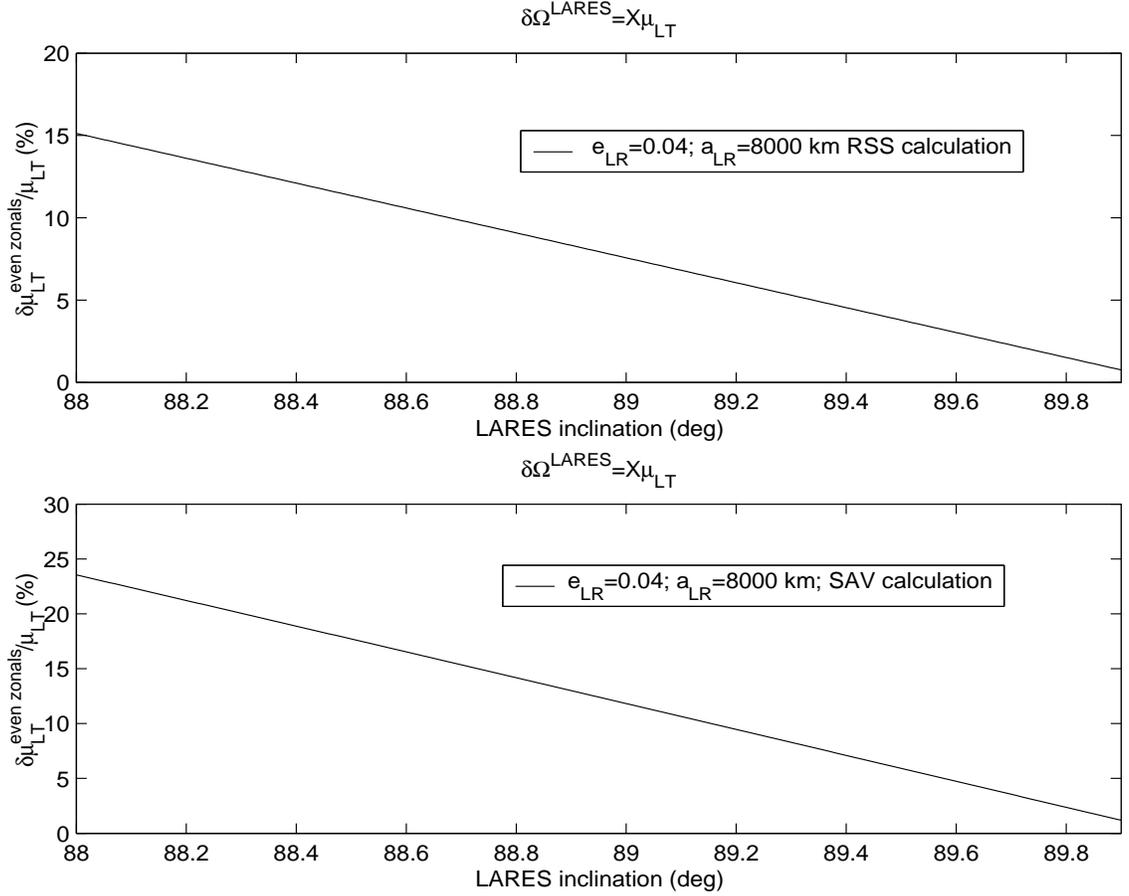}
\end{center}
\caption{\label{solonodo_i} Dependence on $i$ of the percent
1-sigma systematic error $\delta\mu^{\rm even\ zonals}_{\rm
LT}/\mu_{\rm LT}$ due to the mismodelling in the static part of
the even zonal harmonics of the geopotential according to
EIGEN-CG01C for the residuals of the node $\delta\Omega^{\rm
LARES}$ of a nearly-polar orbital configuration with $a=8000$ km
and $e=0.04$. The secular variations of the even zonal harmonics
$\dot J_2,\dot J_4,\dot J_6$ has not been included. In the upper
panel the Root-Sum-Square (RSS) calculation is presented, while in
the lower panel the linear Sum of the Absolute Values (SAV) of the
individual errors is shown.}
\end{figure}

\lb{polare}
The 1-sigma error due to the geopotential, according to
EIGEN-CG01C, for a nearly-polar orbital geometry is depicted in
Figure \ref{solonodo_i}.
For $i=89.9$ deg the systematic error due to the geopotential
amounts to 1-2$\%$; unfortunately, the period of the node, which
is also the period of the gravitational perturbation due to the
solar $K_1$ tide, amounts to $\sim 10^4$ days. As already shown in
(Peterson 1997, Iorio 2002a, 2005b), the mismodelled part of such
a tidal perturbation would act as a huge superimposed bias on the
Lense-Thirring trend over a reasonable observational time span of
some years. On the other hand, for $i=88$ deg the node period
would reduce to $\sim 10^3$ days but the systematic error due to
the mismodelling in the static part of geopotential would amount
to 15-25$\%$.

A further source of uncertainty is also represented by the impact
of the secular variations $\dot J_{\ell}$ of the low-degree even
zonal harmonics whose shift on the node grows quadratically in
time.

\subsubsection{The non-polar scenario} 
\begin{figure}
\begin{center}
\includegraphics[width=15cm,height=12cm]{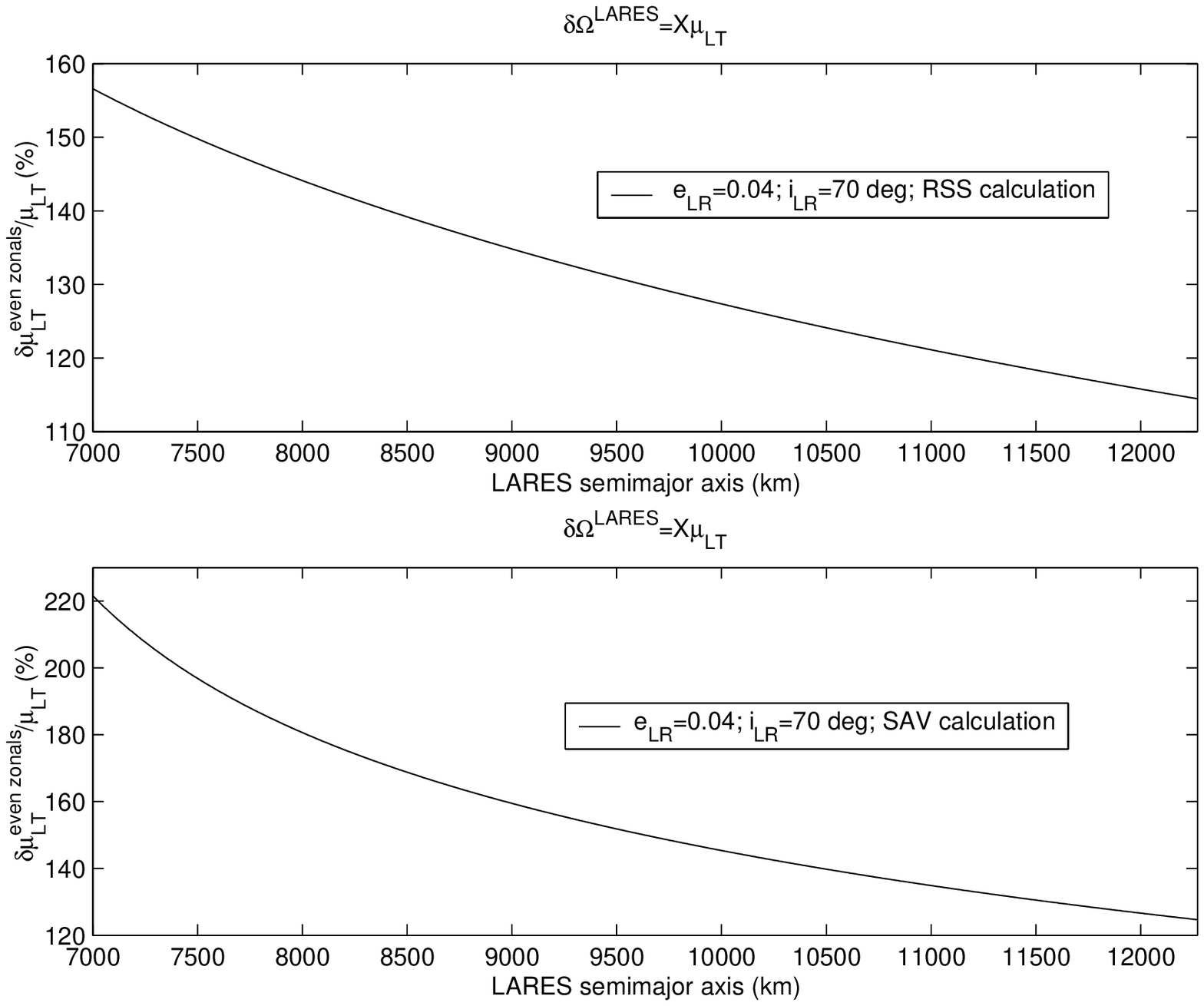}
\end{center}
\caption{\label{solonodo_a} Dependence on $a$ of the percent
1-sigma systematic error $\delta\mu^{\rm even\ zonals}_{\rm
LT}/\mu_{\rm LT}$ due to the mismodelling in the static part of
the even zonal harmonics of the geopotential according to
EIGEN-CG01C for the residuals of the node $\delta\Omega^{\rm
LARES}$ of a non-polar orbital configuration with $i=70$ deg and
$e=0.04$. The secular variations of the even zonal harmonics $\dot
J_2,\dot J_4,\dot J_6$ has not been included. In the upper panel
the Root-Sum-Square (RSS) calculation is presented, while in the
lower panel the linear Sum of the Absolute Values (SAV) of the
individual errors is shown.}
\end{figure}

In order to avoid the
problems with the $K_1$ tide an inclination far from 90 deg could,
in principle, be adopted. Indeed, for $i=70$ deg the period of the
node would amount to $\sim 10^2$ days. However, for non
polar-orbital geometries the mismodelling in the static part of
the geopotential would be fatal independently of the chosen
semimajor axis, as depicted in Figure \ref{solonodo_a}.
The secular variations $\dot J_{\ell}$ of the low degree even
zonal harmonics would further corrupt the measurement of the
Lense-Thirring effect.
\subsubsection{Conclusions}
These considerations clearly show that the possibility of only
analyzing the node of a suitably built passive satellite is not
viable because of the impact of all the uncancelled even zonal
harmonics of the geopotential along with their secular variations.
The polar geometry would allow to reduce the consequences of these
sources of error but, on the other hand, it would enhance the
aliasing effect of the tidal perturbations due to the $K_1$ tide
whose period is equal to that of the node which would amount to
$10^3-10^4$ days.
\subsection{A high-altitude LARES scenario}
For the sake of completeness, we examine the possibility of
adopting for the new satellite a semimajor axis much larger than
that of LAGEOS. E.g., a value $a=36000$ km has been suggested in
(Ciufolini 2004). Apart from the fact that such a mission would
become even more expensive than the originally proposed
LAGEOS-LARES scenario, it would be unfeasible because of the
geopotential's mismodelling.

Indeed, the node of such a high-altitude satellite could not be
combined with those of LAGEOS and LAGEOS II because it would be
weighted by a quite large coefficient which would enhance the
time-dependent perturbations of gravitational and
non-gravitational origin. Among the tidal perturbations, that
induced by the $K_1$ tide would be fatal because its period, equal
to that of the satellite's node, would be $\sim 10^4$ days long.
Such issues have been addressed in (Iorio 2002b, Vespe and
Rutigliano 2004).

Also the possibility of using only the node is not viable. Indeed,
for such an altitude the Lense-Thirring precession is $\sim$ 1 mas
yr$^{-1}$. For $i=70$ deg  and $e=0.04$ the sum of the mismodelled
precessions induced by the even zonal harmonics would be $\sim
100\%$ of the gravitomagnetic effect mainly due to $J_2$ and
$J_4$, according to EIGEN-CG01C. For $i=89$ deg and $e=0.04$ the
impact of the static part of the geopotential would be reduced to
$5\%$ of the Lense-Thirring effect, but the $K_1$ tide would
induce an aliasing semisecular perturbation with a period of
$10^5$ days.
\subsection{The linear combination scenario with LAGEOS and LAGEOS
II}
\begin{figure}
\begin{center}
\includegraphics[width=15cm,height=12cm]{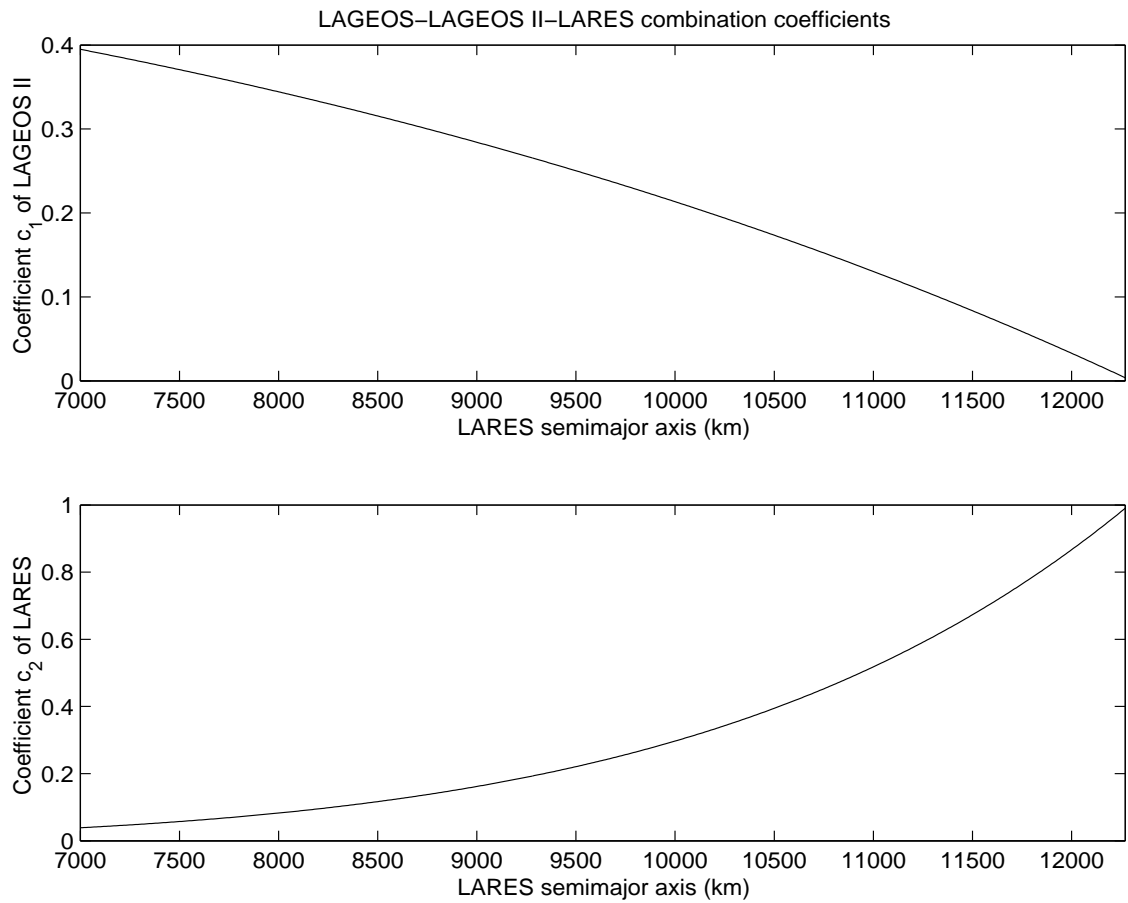}
\end{center}
\caption{\label{COEFF} Dependence on $a$ of the coefficients $c_1$
and $c_2$ with which the nodes of LAGEOS II and LARES enter the
combination of \rfr{tricomb}. The values $i=70$ deg and $e=0.04$
have been assumed.}
\end{figure}

\begin{figure}
\begin{center}
\includegraphics[width=15cm,height=12cm]{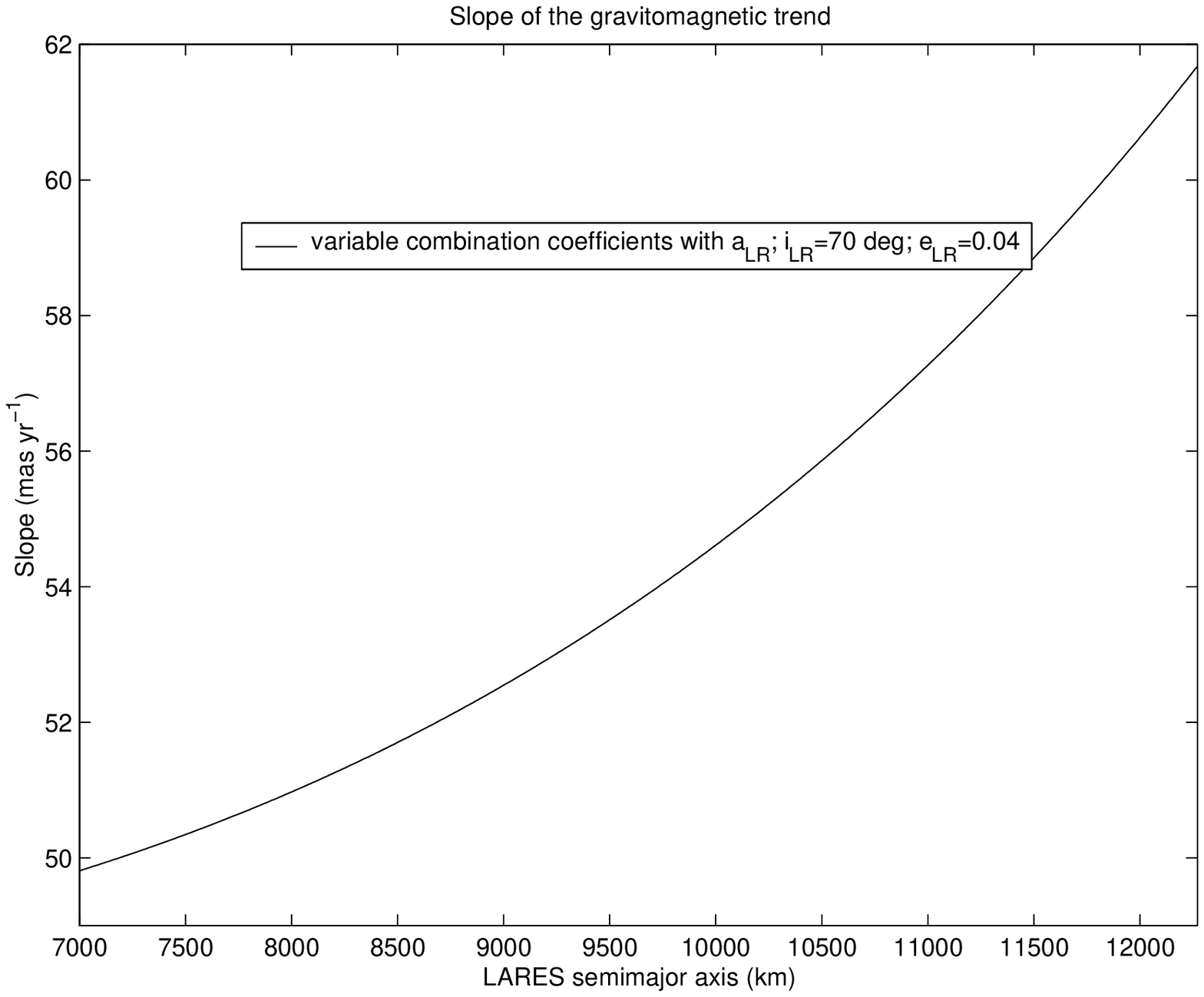}
\end{center}
\caption{\label{trend} Dependence on $a$ of the slope $X_{\rm
LT}$, in mas yr$^{-1}$, of the gravitomagnetic trend of the
combination \rfr{tricomb}. }
\end{figure}

\lb{Kombi} The most viable approach with only one new dedicated
passive satellite at our disposal would be the suitable
combination of its data with those of the already existing LAGEOS
and LAGEOS II. Such proposal was already put forth in  (Iorio et
al. 2004) in the framework of the relativity-dedicated OPTIS
mission (L\"{a}mmerzahl et al. 2004) by retaining for the new
satellite the nominal configuration of LARES. A previous analysis
involving also the perigee of LARES and EGM96 can be found in
(Iorio et al. 2002). According to a well established procedure
(Ciufolini 1996; Iorio and Morea 2004, Iorio 2004, Vespe and
Rutigliano 2004), it is possible to design a linear combination of
the residuals of the nodal rates of LAGEOS, LAGEOS II and LARES
\eqi\delta\dot\Omega^{\rm LAGEOS}+c_1\delta\dot\Omega^{\rm LAGEOS
II}+c_2\delta\dot\Omega^{\rm LARES}\sim X_{\rm LT}\mu_{\rm
LT}\lb{tricomb}\eqf which would cancel out the first two even
zonal harmonics of the geopotential along with their temporal
variations.  The parameter $\mu_{\rm LT}$  is 0 in the Newtonian
mechanics and 1 in the General Theory of Relativity. The
coefficients can be found in Figure \ref{COEFF} and the slope of
the gravitomagnetic trend in Figure \ref{trend}.
If, on the one hand, the inclusion of the existing LAGEOS
satellites would prevent to go below the $1\%$ level due to the
non-gravitational systematic errors, on the other hand such goal
could be reached with a much less expensive mission because, as we
will show, the constraints on the new satellite's orbit could be
greatly weakened thanks to the new Earth gravity models from CHAMP
and GRACE. It would be possible to choose a much smaller semimajor
axis and also the inclination could safely range a wider span near
70 deg without sensibly affecting the systematic error due to the
uncancelled even zonal harmonics of the geopotential. Another
important advantage is that the secular variations $\dot J_2,\dot
J_4$ would not affect this combination.

\begin{figure}
\begin{center}
\includegraphics[width=15cm,height=12cm]{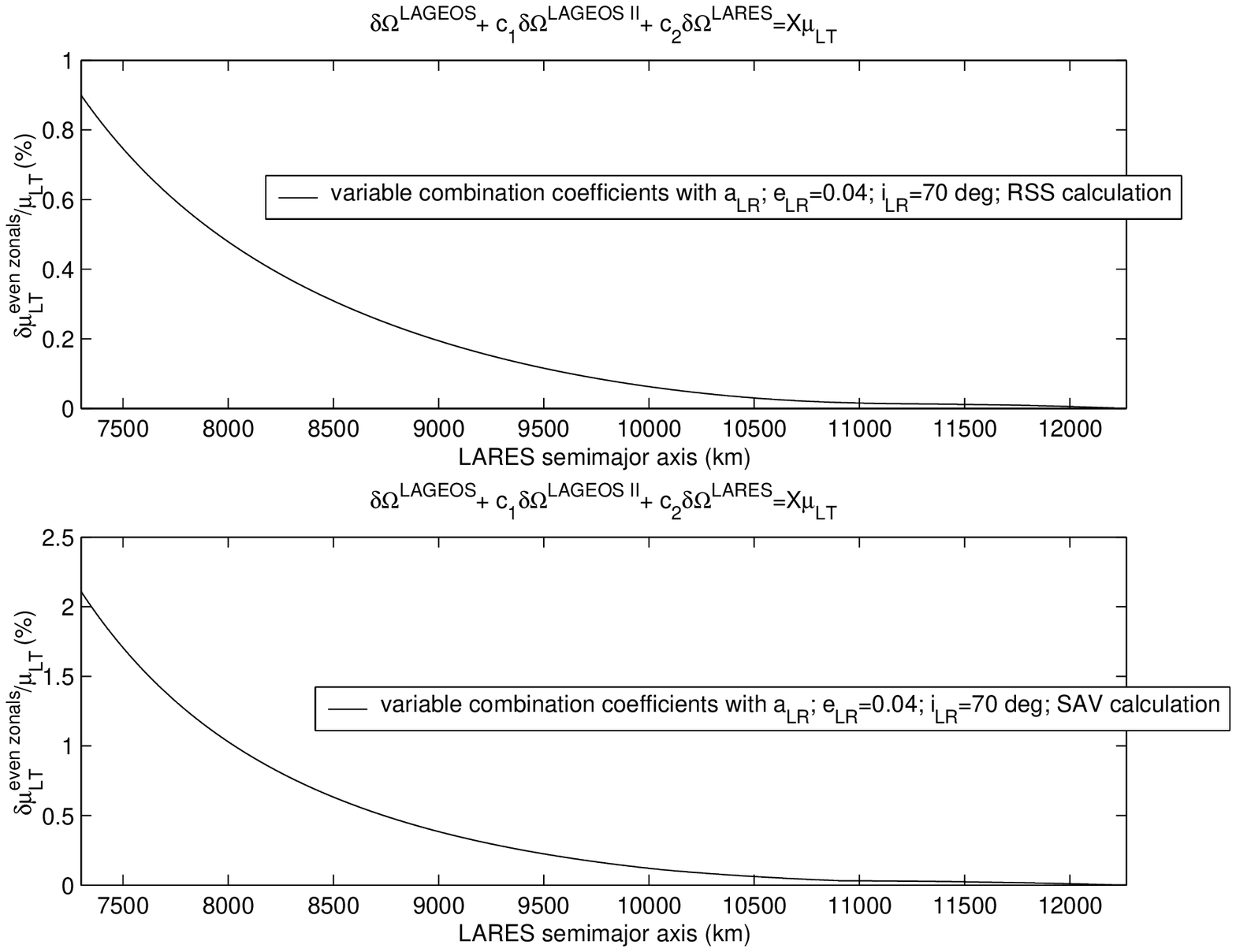}
\end{center}
\caption{\label{L_L2_LR_a_70_CG01C} Dependence on $a$ of the
percent 1-sigma systematic error $\delta\mu^{\rm even\
zonals}_{\rm LT}/\mu_{\rm LT}$ due to the mismodelling in the
static part of the even zonal harmonics of the geopotential
according to EIGEN-CG01C for a linear combination
$\delta\Omega^{\rm LAGEOS}+c_1\delta\Omega^{\rm LAGEOS\
II}+c_2\delta\Omega^{\rm LARES}$ with the existing LAGEOS and
LAGEOS II with $i_{\rm LR }=70$ deg. In the upper panel the
Root-Sum-Square (RSS) calculation is presented, while in the lower
panel the linear Sum of the Absolute Values (SAV) of the
individual errors is shown.}
\end{figure}

In Figure \ref{L_L2_LR_a_70_CG01C} the dependence of the
systematic error due to the geopotential on the semimajor axis is
presented.
It is important to note that even with a semimajor axis of
7500-8000 km the present-day mismodelling in the even zonal
harmonics of the geopotential the $1\%$ level in their impact
could be easily reached. Note also that for such an altitude a
calculation of the systematic error of gravitational origin up to
degree $\ell=20$ is well adequate and reliable: indeed, it can be
shown that the result does not change by adding further terms of
higher degree. The forthcoming more robust and reliable Earth
gravity models from CHAMP and GRACE should further improve this
situation. Moreover, no semisecular time-dependent perturbations
would be introduced: indeed, the period of the node of LARES would
amount to $\sim 10^2$ days for $a=7500-8000$ km.

\begin{figure}
\begin{center}
\includegraphics[width=15cm,height=12cm]{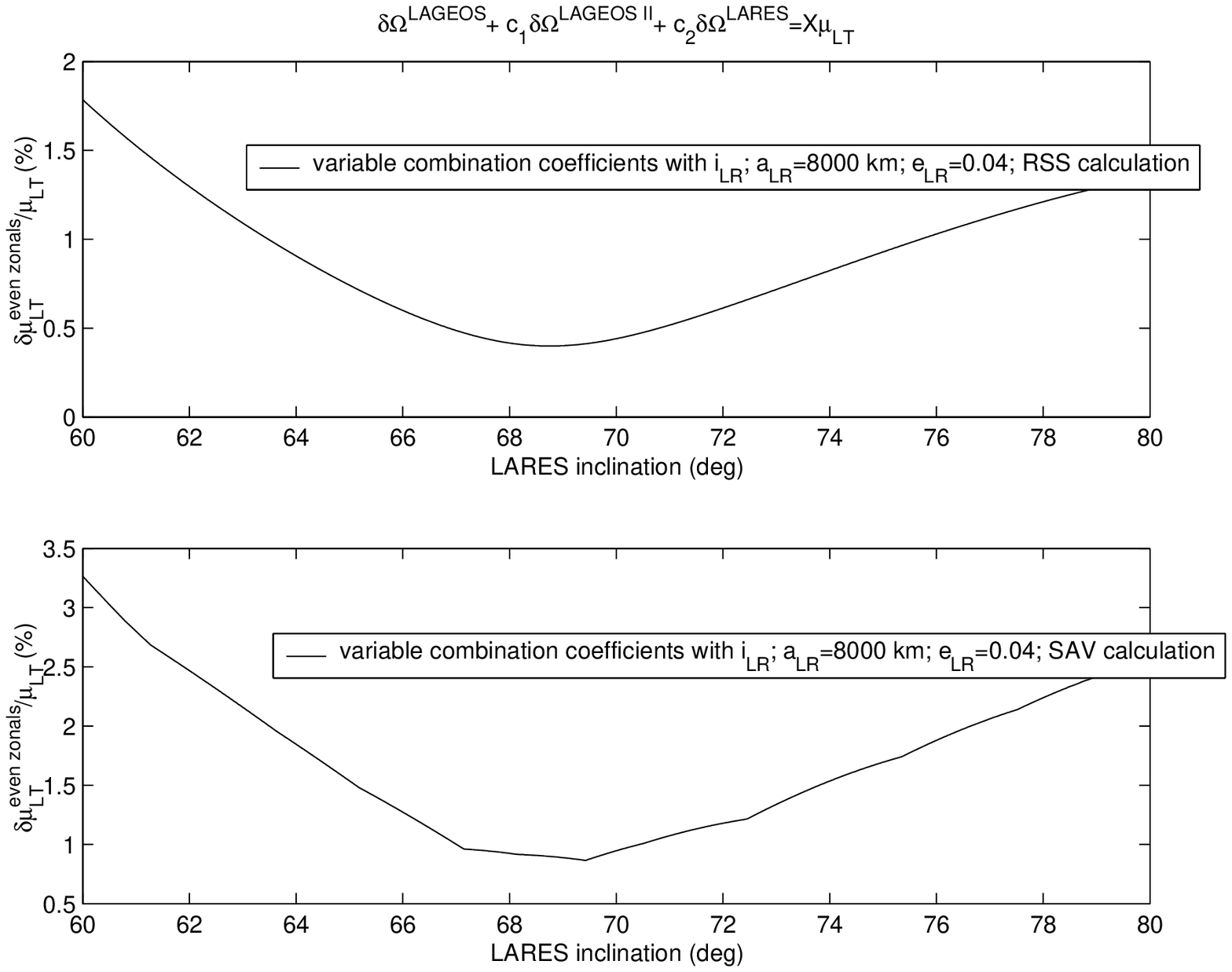}
\end{center}
\caption{\label{INKLI} Dependence on $i$ of the percent 1-sigma
systematic error $\delta\mu^{\rm even\ zonals}_{\rm LT}/\mu_{\rm
LT}$ due to the mismodelling in the static part of the even zonal
harmonics of the geopotential according to EIGEN-CG01C for a
linear combination $\delta\Omega^{\rm LAGEOS}+c_1\delta\Omega^{\rm
LAGEOS\ II}+c_2\delta\Omega^{\rm LARES}$ with the existing LAGEOS
and LAGEOS II with $a_{\rm LR }=8000$ km. In the upper panel the
Root-Sum-Square (RSS) calculation is presented, while in the lower
panel the linear Sum of the Absolute Values (SAV) of the
individual errors is shown.}
\end{figure}

In regard to the departure of the inclination from the nominal
value of 70 deg, Figure \ref{INKLI}, obtained for $a=8000$ km,
clearly shows that the systematic bias of gravitational origin is
rather insensitive to the orbital injection errors over a wide
range.
However, it must be noted that an analysis conducted from 0 deg to
180 deg has revealed that the coefficients $c_1$ of LAGEOS II and
$c_2$ of the new satellite become $\gg 1$ for $i\sim 50$ and 129
deg, while for a polar orbit $c_2$ blows up.

The EUROCKOT facilities at Plesetsk cosmodrome, located at
latitude 62.7 deg N, would easily allow to get the inclination
$i=75.3$ deg for both circular and elliptical orbits in the
altitude range from 200 km up to 2000 km (see Chapter 3 General
Performance Capabilities of the EUROCKOT User Guide
http://www.eurockot.com/pb-pic/20041084.pdf).

Note that Figure \ref{COEFF} yields another possible reason to
further reduce the cost of the mission: indeed, the coefficient
$c_2$ with which the node of LARES enters the combination of
\rfr{tricomb} becomes smaller than 0.1 for low altitudes. This
means that, if one is only interested in the node for the
Lense-Thirring effect, it would not be necessary a particular
technological and financial effort in building the satellite in
order to reduce the action of the non-gravitational perturbations
with respect to the level of the existing LAGEOS satellites.

Finally, one may ask if the combination of \rfr{tricomb} could be
implemented with some of the other existing laser-tracked
satellites: after all, the majority of them orbit at altitudes
much lower than LAGEOS and LAGEOS II and for some of them the
centimeter accuracy in reconstructing their orbits has been
obtained. The most suitable candidate seems to be Ajisai, whose
orbital parameters are $a=7870$ km, $e=0.001$ and $i=50$ deg.
Unfortunately, it turns out that its inclination lies in a domain
in which the systematic error due to the even zonal harmonics
would be unacceptably large: indeed, it would amount to $\sim
20\%$ according to EIGEN-CG01C.

\section{What can be done with two new spacecraft?}\lb{dua}
The possibility of using more than one new passive laser-ranged
satellite together with the existing LAGEOS and LAGEOS II was
investigated for the first time in (Peterson 1997).

In (Iorio 2003b, 2003c, Iorio and Lucchesi 2003) it has been shown
that for a pair of satellites in supplementary orbital
configuration also the differences of the perigees could be used
in addition to the sum of the nodes. Indeed, the Lense-Thirring
perigee precessions depend on $\cos i$, while the aliasing
classical precessions induced by the even zonal harmonics of the
geopotential are of the form (Iorio 2003d)
\eqi\dot\omega^{(J_{\ell})}=A^{(\ell)}\cos^2 i
\left[\sum^{\ell}_{k=2}p^{(\ell)}_k\sin^{k}i +B^{(\ell)}
\right]+C^{(\ell)}\left[\sum^{\ell}_{h=2}q^{(\ell)}_h\sin^{h}i\right]+D^{(\ell)}\eqf
with $h,k$ even, $p^{(\ell)}_k, q^{(\ell)}_k$ numerical
coefficients which depend only on the degree $\ell$ and
$A^{(\ell)},B^{(\ell)},C^{(\ell)},D^{(\ell)}$ different functions
of $a$ and $e$ for the various degrees $\ell$. Then, with the
differences of the perigees for identical orbital configurations
and supplementary inclinations the gravitomagnetic precessions add
up and all the classical precessions cancel out. Such observable
could not be implemented by using many of the already existing SLR
satellites due to the small eccentricities of their orbits: in
(Iorio 2003b) it has been shown that LAGEOS II would not be
suitable for implementing the difference of the perigee because of
certain long-period perturbations of gravitational and
non-gravitational origin which affect its perigee\footnote{In
(Peterson 1997) it has also been shown that neither the sum of the
nodes would be a viable opportunity if LAGEOS II was one of the
partners of the supplementary configuration.}. In (Iorio and
Lucchesi 2003) a configuration of two new passive satellites of
LAGEOS-type in eccentric orbits has been investigated: it turned
out that the difference of the perigees would be affected at a
$\sim 5\%$ level, mainly due to the impact of the
non-gravitational perturbations. A $\sim 1\%$ level could be
achievable with the sum of the nodes, but it is clear that the
implementation of the supplementary orbital configuration with a
pair of passive spacecraft would not be competitive with that
previously analyzed which would allow to reach the same goal--a
$\sim 1\%$ measurement of the Lense-Thirring effect--with just one
new satellite.

Instead, if an active drag-free technology was used, the
implementation of the supplementary configuration with two
satellites would be very interesting. Indeed, both the nodes and
the perigees would be available and it would be possible to
analyze the sum of the nodes, the difference of the perigees and
some suitable linear combination of them without resorting to the
existing LAGEOS and LAGEOS II. In this way the Lense-Thirring
effect could be measured by reducing the total systematic error
below the $1\%$ level and fully exploiting the benefits of the
CHAMP and GRACE gravity models.

Of course, the cost of such an option would be much higher than
those required to built just one passive laser target and insert
it into a relatively low orbit.

\section{Conclusions}
The notable improvements in our knowledge of the classical part of
the terrestrial gravitational potential due to the dedicated CHAMP
and, especially, GRACE missions would make  the launch of a third
satellite of LAGEOS-type dedicated to the measurement of the
Lense-Thirring effect cheaper than it has been considered until
now. Indeed, while the originally proposed scenario involving the
analysis of the sum of the nodes of LAGEOS and LARES would be
relatively expensive to be implemented due to the need of
obtaining the same high altitude of LAGEOS, it would, instead, be
possible to suitably combine the data of the new satellite with
those of the existing LAGEOS and LAGEOS II in order to cancel out
$J_2$ and $J_4$ along with their temporal variations abandoning
the originally proposed stringent requirements on the orbital
geometry of the new laser target. In particular, it would be
possible to reduce the semimajor axis$-$and, then, the cost of the
launcher$-$from that of LAGEOS, i.e. 12270 km, down to 7500-8000
km keeping the systematic error due to the mismodelled even zonal
harmonics of the geopotential within the $1\%$ level (at 1-sigma)
according to the EIGEN-CG01C Earth gravity model. Although the
optimal inclination still remains close to 70 deg, departures of
many degrees from such a value are now well acceptable. For a
combination with LAGEOS and LAGEOS II the only forbidden windows
are near 50, 90 and 130 deg. Moreover, no semisecular long-period
perturbations would be introduced because the period of the node
of the new satellite would be of the order of $10^2$ days.
Finally, with the low-altitude option it would also be possible to
save money on the construction of the satellite because it would
not be necessary to greatly reduce the impact of the
non-gravitational perturbations with respect to the level of the
existing LAGEOS satellites thanks to the small value ($\leq 0.1$)
of the coefficient with which the new object would enter the
combination. In this way the realization of a cheap space mission
dedicated to the measurement of the Lense-Thirring effect with a
$\sim 1\%$ level of accuracy becomes much easier, at least in
principle. Such  a precision is better than what could be obtained
by reanalyzing the data of the existing satellites. Moreover, the
problems with the secular variations of the even zonal harmonics
which affect the performed test with the LAGEOS-LAGEOS II two-node
combination would greatly be reduced. On the other hand, one could
legitimately ask if such an expense is justified for a satellite
whose principal (or unique) goal is the measurement of the
Lense-Thirring effect, while with the already performed or
proposed tests with the existing satellites the obtainable
accuracy in testing the gravitomagnetic effect on the orbit of a
test particle is only ten-twenty times worse. If, instead, it was
possible to also include this goal in the list of the scientific
objectives of other already planned or proposed space missions,
like e.g. OPTIS, without notably increasing their costs or even
suggesting ways to reduce them, this would be a major achievement.
E.g. it seems that a new geodetic satellite at a low-medium
altitude might be useful for many geodetic and geodynamic purposes
like a better knowledge of the centre-of-mass of our planet (Ries
2005, private communication). In regard to OPTIS, the minimum
orbital height should be $\sim 1000$ km, i.e. $a\sim 7378$ km, due
to the limitations posed by the drag-free technology to be adopted
based on the field emission electrical propulsion (FEEP): this
requirement would be compatible with the lower altitude scenario
outlined in the paper.

Since it would not be possible to use the data only from such a
new satellite, the fact that they should be combined with those
from LAGEOS and LAGEOS II sets the limit of the obtainable
systematic error to $\sim 1\%$ due to the action of the
non-gravitational perturbations on the existing twins,
irrespectively of the fact that the gravitational error could be
reduced well below such a cutoff. Only the launch of at least two
new satellites endowed with some active mechanism of compensation
of the non-conservative accelerations and in rather elliptic
orbits would allow to discard the existing LAGEOS satellites fully
exploiting the possibilities opened by the new Earth gravity
models from CHAMP and GRACE and pushing the total systematic error
below the $1\%$ level.

In conclusion, a reliable $1\%$ measurement of the Lense-Thirring
effect could be obtained with a new satellite placed in a
relatively low orbit ($a=7500-8000$ km) which would allow to
greatly reduce the costs with respect to, e.g., the originally
proposed LARES mission. The node of such a new satellite, which
could also be passive and built without resorting to any
particularly sophisticated technologies, should be combined with
those of LAGEOS and LAGEOS II in order to cancel out the first two
even zonal harmonics along with their secular variations. The
limit of the reduction of the systematic error is set by the
non-gravitational perturbations acting on the existing twin
satellites, irrespectively of the improvements allowed by the new
forthcoming CHAMP/GRACE-based Earth gravity models. A total
systematic error smaller than $1\%$ could be obtained with at
least two new drag-free spacecraft.
\section*{Acknowledgments}
I am grateful to J. Ries for useful discussions.


\end{document}